\documentclass[aps,prl,twocolumn,superscriptaddress,showpacs]{revtex4}
\usepackage{graphicx}
\begin{document}


\title{Absence of Jahn-Teller transition in the hexagonal Ba$_3$CuSb$_2$O$_9$ single crystal}

\author{N. Katayama}
\affiliation{Department of Applied Physics, Nagoya University, Nagoya 464-8603, Japan}
\author{K. Kimura}					
\thanks{Present address: Division of Materials Physics, Graduate School of Engineering Science, Osaka University, Toyonaka 560-8531, Japan}
\affiliation{Institute for Solid State Physics, University of Tokyo, Kashiwa 277-8581, Japan}
\author{Y. Han}					
\affiliation{KYOKUGEN (Center for Quantum Science and Technology under Extreme Conditions), Osaka University, Toyonaka, Osaka 560-8531, Japan}
\affiliation{Wuhan National High Magnetic Field Center, Huzhong University of Science and Technology, Wuhan 430074, China}
\author{J. Nasu}					
\affiliation{Department of Physics, Tohoku University, Sendai 980-8578, Japan}
\author{N. Drichko}					
\affiliation{Institute for Quantum Matter (IQM) and Department of Physics and Astronomy, Johns Hopkins University, Baltimore, MD 21218, USA}
\author{Y. Nakanishi}					
\affiliation{Graduate School of Engineering Iwate University Morioka 020-8551, Japan}
\author{M. Halim}					
\affiliation{Institute for Solid State Physics, University of Tokyo, Kashiwa 277-8581, Japan}
\author{Y. Ishiguro}					
\affiliation{Division of Materials Physics, Graduate School of Engineering Science, Osaka University, Toyonaka 560-8531, Japan}
\author{R. Satake}
\affiliation{Department of Applied Physics, Nagoya University, Nagoya 464-8603, Japan}
\author{E. Nishibori}
\thanks{Present address: RIKEN SPring-8 Center, Kouto, Sayo, Hyogo 679-5148, Japan}
\affiliation{Department of Applied Physics, Nagoya University, Nagoya 464-8603, Japan}
\author{M. Yoshizawa}					
\affiliation{Graduate School of Engineering Iwate University Morioka 020-8551, Japan}
\author{T. Nakano}					
\affiliation{Department of Physics, Graduate School of Science, Osaka University, Toyonaka, Osaka 560-0043, Japan}
\author{Y. Nozue}					
\affiliation{Department of Physics, Graduate School of Science, Osaka University, Toyonaka, Osaka 560-0043, Japan}
\author{Y. Wakabayashi}					
\affiliation{Division of Materials Physics, Graduate School of Engineering Science, Osaka University, Toyonaka 560-8531, Japan}
\author{S. Ishihara}					
\affiliation{Department of Physics, Tohoku University, Sendai 980-8578, Japan}
\author{M. Hagiwara}					
\affiliation{KYOKUGEN (Center for Quantum Science and Technology under Extreme Conditions), Osaka University, Toyonaka, Osaka 560-8531, Japan}
%
%
\author{H. Sawa}					
\affiliation{Department of Applied Physics, Nagoya University, Nagoya 464-8603, Japan}
\author{S. Nakatsuji}					
\affiliation{Institute for Solid State Physics, University of Tokyo, Kashiwa 277-8581, Japan}
\affiliation{PRESTO, Japan Science and Technology Agency (JST), 4-1-8 Honcho Kawaguchi, Saitama 332-0012, Japan}
\date{\today}

\begin{abstract}
We present a comprehensive structural study on perovskite-type 6$H$-Ba$_3$CuSb$_2$O$_9$, which exhibits a spin-orbital short-range ordering on a honeycomb-based lattice.  By combining synchrotron x-ray diffraction, electron spin resonance, ultrasound measurement and Raman spectroscopy, we found that the static Jahn-Teller distortion is absent down to the lowest temperature in the present material, indicating orbital ordering is strongly suppressed. We discuss such an unusual state is realized with the help of spin degree of freedom, leading to a spin-orbital entangled liquid state.
\end{abstract}

\maketitle


Spin liquids have been widely recognized as a new state of matter, as an increasing number of candidates with quantum spin $S$ = 1/2 have been found recently \cite{rf:5,rf:16,rf:17,rf:8}. On the other hand, quantum liquids based on another electronic degree of freedom, orbital, has been believed unrealistic, because the energy of orbital correlation is normally one order of magnitude stronger than spin-exchange coupling, leading to an orbital ordering at a significantly high temperature accompanied by a cooperative Jahn-Teller (JT) distortion. However, if we may bring down the orbital energy to the same scale, it might lead to a novel ``spin-orbital liquid" state. While some candidates have been proposed through extensive experimental and theoretical studies \cite{rf:30,rf:40,rf:31}, they were later found out to have orbital freezing at low temperature ($T_s$) \cite{rf:32,rf:33,rf:34}. Thus, an experimental realization of such spin-orbital liquid has remained challenge.


\begin{figure}
\includegraphics[width=7.9cm]{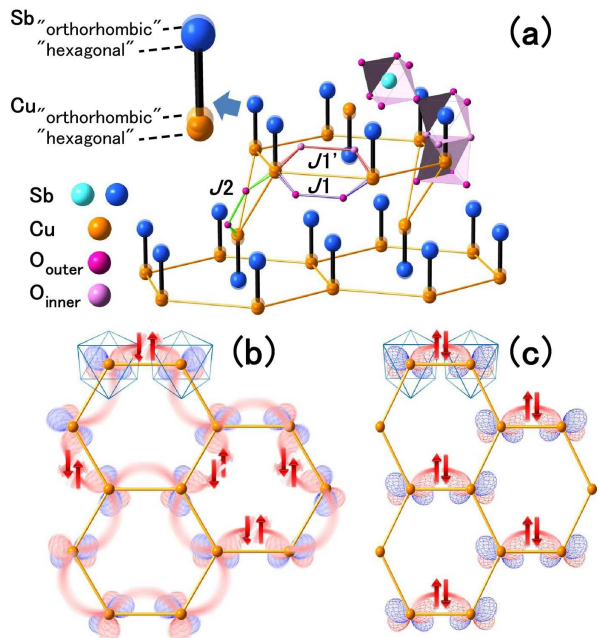}
\caption{\label{fig:Fig4}
(color online) (a) Schematic view of the local structure for hexagonal and orthorhombic samples. (b)(c) Schematics of spin-singlet formation in short-range honeycomb lattices of Cu$^{2+}$ for (b) hexagonal and (c) orthorhombic samples. For (b), a spin-orbital entangled short-range-order state is expected. A pair of up and down arrows indicates a singlet state of the dimer based on the neighboring Cu$^{2+}$ spins. At each site, an unpaired electron of Cu$^{2+}$ occupies the $d_{x^2-y^2}$, $d_{y^2-z^2}$ or $d_{z^2-x^2}$ orbital.}
\end{figure}

Perovskite-type 6$H$-Ba$_3$CuSb$_2$O$_9$ may provide a promising opportunity to explore the novel spin-orbital liquid state. Recently, we reported that spin-orbital short-range ordering occurs in the short-range honeycomb lattice of Cu$^{2+}$, as depicted in Fig.\ref{fig:Fig4}(a) \cite{rf:11}, sharply contradicts to triangular lattice system as was reported previously \cite{rf:20}. In addition to the confirmation of a dynamic spin state down to 20 mK by $\mu$SR \cite{rf:11, rf:21, rf:22}, powder x-ray diffraction clearly indicate that even at low temperature, the hexagonal components remain along with some orthorhombically distorted components. In the hexagonal phase, three-fold symmetry exists for the Cu$^{2+}$ sites, indicating the absence of a cooperative JT distortion. To explain this unusual feature, we proposed two possible scenarios. (i) A $non$-cooperative static JT distortion appears. In this scenario, the local symmetry is lowered by a static JT distortion, but the overall hexagonal symmetry remains. (ii) The static JT distortion is absent and instead, a dynamic JT distortion appears, leading to a novel spin-orbital liquid state, as depicted in Fig.\ref{fig:Fig4}(b). These two possible scenarios cannot be distinguished by experimental results using powder specimens alone; a thorough structural study is required using a single crystal without orthorhombic components. Here, we report the comprehensive study on a hexagonal single crystal of Ba$_3$CuSb$_2$O$_9$ that exhibits no cooperative JT transition down to low temperatures. This provides the first example of a copper 3$d^9$ compound with no JT transition. Our results suggest a formation of a spin-orbital liquid state.

Single crystals of Ba$_3$CuSb$_2$O$_9$ were grown from a BaCl$_2$ flux under oxygen gas flow at ISSP, University of Tokyo. A mixture of polycrystalline Ba$_3$CuSb$_2$O$_9$ and BaCl$_2$ was inserted into a Pt crucible and heated to 1500 K, followed by slow cooling. Single-crystal and powder x-ray diffraction experiments were performed at SPring-8 BL02B1 and BL02B2, respectively. For single-crystal x-ray diffraction, a typical size of 40 $\times$ 40 $\times$ 20 $\mu$m$^3$ were measured with the wavelength of 0.35 \AA. An X-band ESR apparatus (Bruker EMX EPR Spectrometer) was used for precise measurements of the temperature and angular dependencies of ESR spectra at about 9.3 GHz. Raman measurements were done with a Jobin-Yvone T64000 triple monochromator spectrometer with an Olympus microscope. The excitation line was the 514.5 nm line of a Spectra-Physics Ar-Kr laser. The probe was 2 $\mu$m in diameter. For low-temperature measurements, He-flow Janis Microcryostat was employed. Velocity measurements of the ultrasonic transverse and longitudinal wave propagating along the $c$-crystallographic axis were performed with a frequency of 15 and 30 MHz, respectively. 

\begin{figure}[btp]
\includegraphics[width=7.8cm]{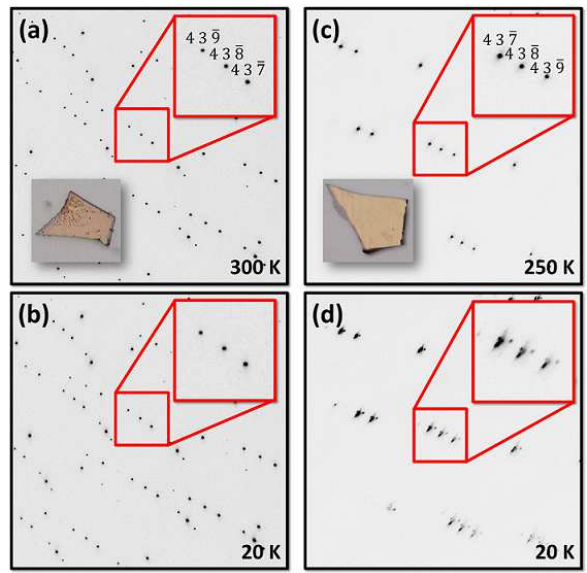}
\caption{\label{fig:Fig1}
(color online) Single crystal x-ray diffraction profiles for (a-b) hexagonal and (c-d) orthorhombic samples. The insets in (a) and (c) are photographs of the transparent brown single crystals for the hexagonal and orthorhombic samples, respectively. The hexagonal samples are darker than the orthorhombic samples.}
\end{figure}

First, let us present the results for the single crystal x-ray diffraction measurements for the sample that retains the hexagonal symmetry down to the lowest temperature. Figures \ref{fig:Fig1}(a) and \ref{fig:Fig1}(b) show profiles obtained at 300 K and 20 K, respectively. The peaks show no signs of splitting or broadening down to 20 K. Hereinafter, we call such a sample a ``hexagonal sample". The hexagonal sample can be well refined by using the centro-symmetric space group $P6_3/mmc$ for all temperatures. For $P6_3/mmc$, the three-fold symmetry is retained for Cu$^{2+}$ sites, clearly indicating the absence of the cooperative JT distortion.

These observations are in sharp contrast with our previous single crystal x-ray diffraction study of 6$H$-Ba$_3$CuSb$_2$O$_9$ \cite{rf:11}. There, we reported that the Bragg peak splits into several separate reflections upon decreasing temperature, as shown in Figs.\ref{fig:Fig1}(c) and \ref{fig:Fig1}(d). This result indicates that the hexagonal $P6_3/mmc$ symmetry is lowered to the orthorhombic $Cmcm$ symmetry. We attribute this effect to a cooperative JT distortion induced by uniform orbital ordering of Cu$^{2+}$ ions (Fig.\ref{fig:Fig4}(c)). Hereinafter, we call such a sample an ``orthorhombic sample". Note that some hexagonal components $\sim$ 1-10 \% of the volume fraction remain in orthorhombic samples, even at the lowest temperature, as previously reported \cite{rf:11}. 

\begin{figure}
\includegraphics[width=8.2cm]{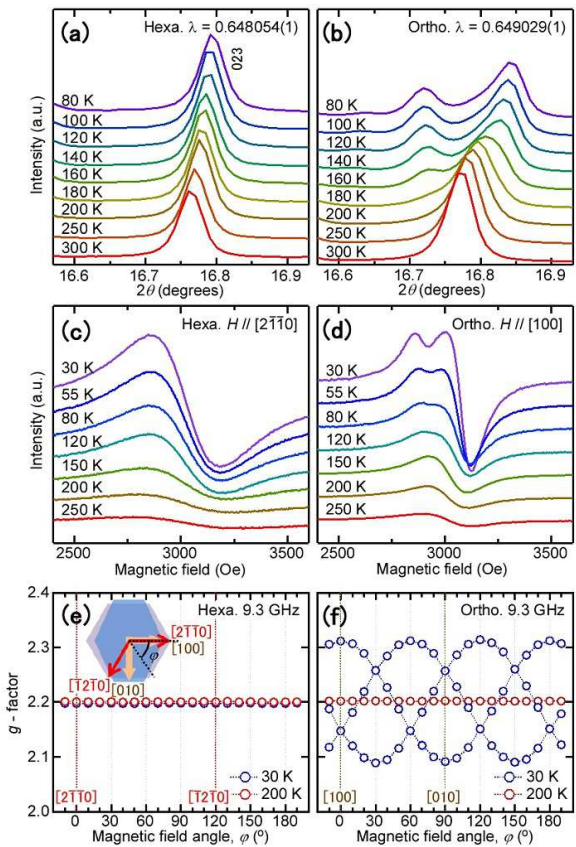}
\caption{\label{fig:Fig2}
(color online) (a)(b) Powder x-ray diffraction patterns of the (a) hexagonal and (b) orthorhombic samples. (c)(d) Temperature-dependent ESR curves along $[2\bar{1}\bar{1}0]$ and $[100]$ of the (c) hexagonal and (d) orthorhombic samples. (e)(f) Angular-dependent $g$ factor in the $c$ plane for hexagonal and orthorhombic samples, respectively. The inset in (e) indicates the Miller indices for the hexagonal (regular hexagon) and orthorhombic samples (distorted hexagon). The rotation angle of the magnetic field relative to the $[2\bar{1}\bar{1}0]$ or $[100]$ axis is given by $\varphi$.}
\end{figure}

The structural features that differentiate between the hexagonal and orthorhombic samples can also be identified by powder x-ray diffraction. We prepared powder samples by crushing the crystals used in the single crystal x-ray diffraction experiments. Below 200 K, no signs of splitting occur for the hexagonal sample down to 80 K, but the peaks split for the orthorhombic sample, as shown in Figs.\ref{fig:Fig2}(a) and \ref{fig:Fig2}(b), respectively. Using a cryostat, we confirmed the hexagonal sample retains hexagonal symmetry down to 13 K (Supplementary Information, Fig.S1). In order to further reveal the unusual state of the hexagonal phase, the local structure associated with the JT distortion should be studied.

Electron spin resonance (ESR) is known to sensitively detect the local orbital configuration through the anisotropy of the $g$ factor \cite{rf:23}. For the magnetic field parallel to the $[2\bar{1}\bar{1}0]$ ([100]) direction of the hexagonal (orthorhombic) sample, we find distinct ESR. For the orthorhombic crystal, the curves split below 200 K (Fig.\ref{fig:Fig2}(d)), corresponding to the hexagonal-orthorhombic structural phase transition found by the x-ray measurement. In contrast, down to 30 K, the hexagonal sample produces the field-derivative of a clear single Lorentzian signal (Fig.\ref{fig:Fig2}(c)). Figures \ref{fig:Fig2}(e) and \ref{fig:Fig2}(f) show the angular dependence of the $g$ factors in the $c$ plane obtained by fitting the ESR curves with the field-derivatives of single- or multiple-peak Lorentzian functions. At 30 K, the orthorhombic sample produces three clear periodic component (Fig.\ref{fig:Fig2}(f)), which originate from the three types of domains corresponding to the three types of the static JT distortion naturally expected for a CuO$_6$ octahedron. In contrast, for the hexagonal phases at all the temperatures in the hexagonal sample as well as above 200 K in the orthorhombic sample, the $g$ factors are almost isotropic along the in-plane directions, as expected for a sample with hexagonal symmetry (Figs.\ref{fig:Fig2}(e) and \ref{fig:Fig2}(f)). This result indicates that the hexagonal sample retains its hexagonal symmetry at temperatures down to 30 K.

\begin{figure}
\includegraphics[width=8.2cm]{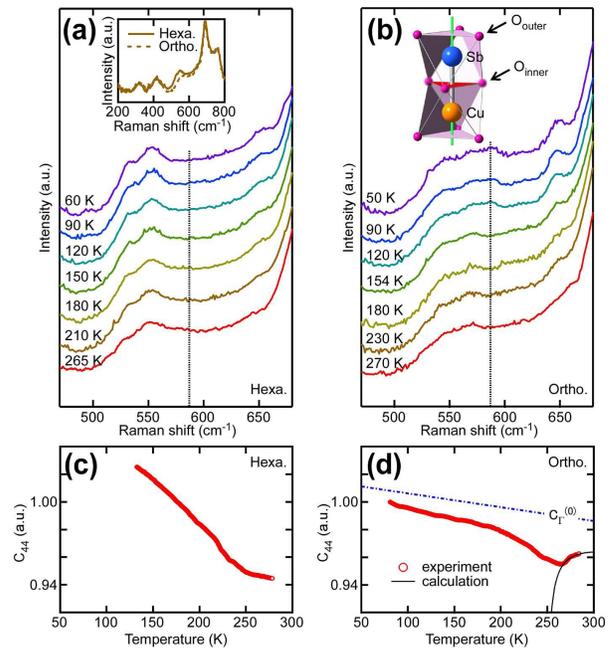}
\caption{\label{fig:Fig5}
(color online) (a)(b) Polarised xx Raman spectra of the (a) hexagonal and (b) orthorhombic samples in the range of oxygen stretching vibrations. Vertical dotted lines are at 587 cm$^{-1}$. Spectra are shifted vertically for clarity. The inset in (a) shows polarised xx Raman spectra in the hexagonal phases of the orthorhombic and hexagonal samples at 240 K. Note the difference in frequencies of the O$_{inner}$ stretching vibrations. The inset in (b) shows the local crystal structure around a JT-active Cu$^{2+}$ ion. (c)(d) Elastic constants of the transverse $C_{44}$ for the (c) hexagonal and (d) orthorhombic samples. The black solid line indicates the theoretical fit with the Curie-Weiss-like formula for the cooperative JT effect.}
\end{figure}

Raman spectroscopy and ultrasound measurements provide other sensitive probes for the local structural symmetry and were used to examine the temperature variation of the JT distortion \cite{rf:41,rf:42,rf:43,rf:44,rf:45}. The overall Raman spectra of Ba$_3$CuSb$_2$O$_9$ are defined by high disorder and broken local symmetry in the edge-sharing octahedra, leading to a split in the shared-face oxygen modes around 550~cm$^{-1}$, as shown in Figs.\ref{fig:Fig5}(a) and \ref{fig:Fig5}(b). The spectra of the orthorhombic sample display new bands at temperatures below the phase transition; for example, the polarised xx Raman spectra shown in Fig.\ref{fig:Fig5}(b) exhibit a new band at 587~cm$^{-1}$ upon cooling, indicating a lowered symmetry. In contrast, the spectra of the hexagonal sample, as shown in Fig.\ref{fig:Fig5}(a), do not change upon lowering the temperature, indicates the symmetry is maintained. Figures \ref{fig:Fig5}(c) and \ref{fig:Fig5}(d) show the the transverse $C_{44}$ elastic constant data collected using ultrasound measurement as a function of temperature. Ultrasound measurement of the orthorhombic sample clearly indicates a characteristic elastic softening in $C_{44}$ towards the phase transition, relevant to the orbitals of $d_{x^2-y^2}$, $d_{y^2-z^2}$ and $d_{z^2-x^2}$ (Fig.\ref{fig:Fig5}(d)). The solid line in Fig.\ref{fig:Fig5}(d) is the theoretical elastic constant based on the following Curie-Weiss-like formula for the cooperative JT effect $C_{\Gamma}$($T$) = $C^{(0)}_{\Gamma}$($T$)($T$ - $T_c$)/ ($T$ - $\Theta$). Here, $C^{(0)}_{\Gamma}$ is a background elastic constant mainly due to the anharmonicity involved in the lattice vibrations. This fitting allows us to estimate the positive paramagnetic Curie-Weiss constant of $\Theta$ $\sim$ 250 K, indicating a ferro-orbital interactions. This is fully consistent with the ferro-type orbital arrangements in orthorhombic samples, as shown in Fig.\ref{fig:Fig4}(c). On the contrary a smooth curve without any characteristic softening is observed down to the lowest temperatures in hexagonal sample, as shown in Fig.\ref{fig:Fig5}(c), again indicating the absence of symmetry lowering. 

Our ESR, Raman spectra and ultrasound measurements are in good accordance with those expected in hexagonal symmetry. While the Raman measurement time scale (psec) is comparable or shorter than a typical dynamic JT time scale \cite{rf:46,rf:47}, the above results of the much slower probes, namely ESR (nsec) and ultrasound measurements ($\mu$sec), clearly exclude the non-cooperative static JT distortion scenario (i) presented above. $Non$-cooperative static JT distortion must break threefold symmetry in each Cu$^{2+}$ site and generate anisotropic $g$ factors along the in-plane directions in ESR experiments, similarly to the low temperature phase of orthorhombic sample, sharply contradicts to that found in the hexagonal sample. It is apparent that orbital degree of freedom exists at high temperature in the present samples because the orthorhombic sample undergoes a cooperative JT transition at around 200 K and furthermore, a dynamic orbital state has been already confirmed at 290 K in a hexagonal sample using an inelastic x-ray scattering experiment \cite{rf:12}. On cooling, the hexagonal sample shows no signs of symmetry lowering, which clearly indicate the absence of the $non$-cooperative static JT distortion in all temperature region. 

In order to further clarify the origin of the absence of the JT transition in the hexagonal sample, we investigate the structural differences between the hexagonal and orthorhombic samples, in particular, in the high-temperature hexagonal phases. To study the local structure around JT-active Cu ions, we refined the coordinates of crystallographically equivalent Cu-Sb sites independently using single-crystal x-ray diffraction. For the refinement, the Cu-Sb composition ratio is fixed to 1.0 for both the hexagonal and orthorhombic samples. Although we detected by chemical analysis \cite{rf:11} a slight departure from unity in the Cu-Sb composition ratio for the orthorhombic sample, this does not significantly affect the parameters obtained (Table S1). Using the bond valence sum technique, we estimated the valences of separated Cu and Sb to be close to 2+ and 5+, respectively, indicating the parameters obtained are reasonable. For the orthorhombic sample, the $z$ coordinates of the face-sharing octahedral sites of Cu and Sb are very similar; however, for the hexagonal sample, the $z$ coordinates of these sites differ. As depicted in Fig.\ref{fig:Fig4}(a), the change in $z$ coordinates can be interpreted as the Cu-Sb dumbbell moving toward the Cu and slightly elongating. Note that even with such a shift, the Cu-Sb dumbbell remains along with the three-fold axis for both structures, indicating the absence of a static JT distortion. The displacement of the Cu-Sb dumbbell leads to longer Cu-O$_{inner}$ bonds and shorter Cu-O$_{outer}$ bonds (Fig.\ref{fig:Fig4}(a)). In the high temperature hexagonal phases, the different structure of face-sharing CuSbO$_9$ octahedra for the hexagonal and orthorhombic samples is also detected by Raman spectroscopy (inset of Fig.\ref{fig:Fig5}(a)). The vibrations of the shared-face oxygens shift clearly from 552 and 576 cm$^{-1}$ for the orthorhombic sample to 535 and 552 cm$^{-1}$ for the hexagonal sample, respectively.

By using the local structural parameters presented above, we estimate the change in superexchange interactions between neighbouring Cu ions for the hexagonal and orthorhombic samples. The perturbational exchange processes are classified into the two forms: two holes located on a single Cu ion or on a single O ion in intermediate states. The changes in the exchange constants that arise because of the change in the Cu-O-O angle are almost canceled out in the two processes. On the other hand, for both processes, decreasing the Cu-O bond length  increases the exchange constant. For the $J$1 exchange interaction, depicted in Fig.\ref{fig:Fig4}(a), Cu-O$_{outer}$ bond lengths are shorter in the hexagonal sample compared with the orthorhombic sample, which increases the transfer integral and thereby enhances the $J$1 exchange interactions. Note that the $J$1' exchange interactions through Cu-O$_{inner}$-O$_{inner}$-Cu paths may be ignored because they are much less than those through Cu-O$_{outer}$-O$_{outer}$-Cu paths owing to the long bond lengths. 

From these estimations of the superexchange interactions, we hypothesise that in the hexagonal sample, the increase in exchange interactions is related to the unusual suppression of a static JT distortion. The orbital energy is typically much higher than the spin energy, resulting in a weak correlation between spin and orbital. However, the orbital energy is reduced due to the spatially separated CuO$_6$ octahedra. If the magnetic exchange interactions are enhanced to the point that they are comparable with the orbital energy, the interplay between spin and orbital would destabilise the conventional orbital-ordered state, leading to a novel spin-orbital entangled state. Indeed, such spin-orbital correlation has been already pointed out by using Huang scattering \cite{rf:12}. Furthermore, a recent theory supports such a scenario, showing that in the absence of the static JT distortion, cooperation between the intersite spin-orbital interaction and the on-site dynamic JT effect stabilises a spin-orbital liquid state \cite{rf:13}. Finally, we should note that the dominant control parameter that tunes the ground state between the hexagonal and orthorhombic phases is still unclear experimentally. A theoretical consideration is given in supplemental information \cite{rf:supple}.

To summarize, our experiments thus clarify the absence of a static JT distortion and the related unique structural features in 6$H$-Ba$_3$CuSb$_2$O$_9$. JT-active Cu$^{2+}$ ions, which are connected with JT-inactive Sb$^{5+}$ ions through ligand oxygens, inherently form a short-range honeycomb lattice, destabilizing the conventional static JT distortion. Instead, the orbital degree of freedom is most likely quenched by a dynamic JT distortion. For the orthorhombic sample, the cooperative JT distortion appears followed by a static orbital ordering at $\sim$200 K. On the other hand, no signs of the static JT distortion down to the lowest temperature measured provides evidence of an ``orbital $non$-frozen state", suggesting the formation of a spin-orbital entangled liquid state. 

The authors acknowledge Prof. C. Broholm for valuable discussion. This work was supported by a Grant-in-Aid for Scientific Research (No.~23244074, No.~242440590, No.~25707030, No.~23102702 and No.~24540354), PRESTO of JST, the Global COE Program (Core Research and Engineering of Advanced Materials and Interdisciplinary Education Center for Materials Science)(Grant No.~G10) from the MEXT and DOE grant for The Institute of Quantum Matter DE-FG02-08ER46544. The synchrotron radiation experiments were performed at SPring-8 with the approval of the Japan Synchrotron Radiation Research Institute (JASRI)(Proposals No. 2011B0083/BL02B1 and No.2011B0084/BL02B2). The use of the Materials Design and Characterization Laboratory at ISSP is gratefully acknowledged. 

\end{document}



\title{Supplemental Information for ``Absence of Jahn-Teller transition in the hexagonal Ba$_3$CuSb$_2$O$_9$ single crystal"}

\author{N. Katayama}
\affiliation{Department of Applied Physics, Nagoya University, Nagoya 464-8603, Japan}
\author{K. Kimura}					
\thanks{Present address: Division of Materials Physics, Graduate School of Engineering Science, Osaka University, Toyonaka 560-8531, Japan}
\affiliation{Institute for Solid State Physics, University of Tokyo, Kashiwa 277-8581, Japan}
\author{Y. Han}					
\affiliation{KYOKUGEN (Center for Quantum Science and Technology under Extreme Conditions), Osaka University, Toyonaka, Osaka 560-8531, Japan}
\affiliation{Wuhan National High Magnetic Field Center, Huzhong University of Science and Technology, Wuhan 430074, China}
\author{J. Nasu}					
\affiliation{Department of Physics, Tohoku University, Sendai 980-8578, Japan}
\author{N. Drichko}					
\affiliation{Institute for Quantum Matter (IQM) and Department of Physics and Astronomy, Johns Hopkins University, Baltimore, MD 21218, USA}
\author{Y. Nakanishi}					
\affiliation{Graduate School of Engineering Iwate University Morioka 020-8551, Japan}
\author{Y. Ishiguro}					
\affiliation{Division of Materials Physics, Graduate School of Engineering Science, Osaka University, Toyonaka 560-8531, Japan}
\author{R. Satake}
\affiliation{Department of Applied Physics, Nagoya University, Nagoya 464-8603, Japan}
\author{E. Nishibori}
\thanks{Present address: RIKEN SPring-8 Center, Kouto, Sayo, Hyogo 679-5148, Japan}
\affiliation{Department of Applied Physics, Nagoya University, Nagoya 464-8603, Japan}
\author{M. Yoshizawa}					
\affiliation{Graduate School of Engineering Iwate University Morioka 020-8551, Japan}
\author{T. Nakano}					
\affiliation{Department of Physics, Graduate School of Science, Osaka University, Toyonaka, Osaka 560-0043, Japan}
\author{Y. Nozue}					
\affiliation{Department of Physics, Graduate School of Science, Osaka University, Toyonaka, Osaka 560-0043, Japan}
\author{Y. Wakabayashi}					
\affiliation{Division of Materials Physics, Graduate School of Engineering Science, Osaka University, Toyonaka 560-8531, Japan}
\author{S. Ishihara}					
\affiliation{Department of Physics, Tohoku University, Sendai 980-8578, Japan}
\author{M. Hagiwara}					
\affiliation{KYOKUGEN (Center for Quantum Science and Technology under Extreme Conditions), Osaka University, Toyonaka, Osaka 560-8531, Japan}
%
%
\author{H. Sawa}					
\affiliation{Department of Applied Physics, Nagoya University, Nagoya 464-8603, Japan}
\author{S. Nakatsuji}					
\affiliation{Institute for Solid State Physics, University of Tokyo, Kashiwa 277-8581, Japan}
\affiliation{PRESTO, Japan Science and Technology Agency (JST), 4-1-8 Honcho Kawaguchi, Saitama 332-0012, Japan}
\date{\today}

\maketitle

{\bf Introduction}\\

This supplementary information is provided in support of the main text with details. ${Section~1}$ presents an additional powder x-ray diffraction data and structural parameters obtained from the single crystal x-ray diffraction experiments. ${Section~2}$ provides four circle x-ray diffraction data for hexagonal and orthorhombic samples. ${Section~3}$ supplies the theoretical consideration about the control parameter which tunes the ground state between the hexagonal phase and the orthorhombic one.\\

{\bf 1. Synchrotron x-ray diffraction experiments}\\

\begin{figure}
\includegraphics[width=8cm]{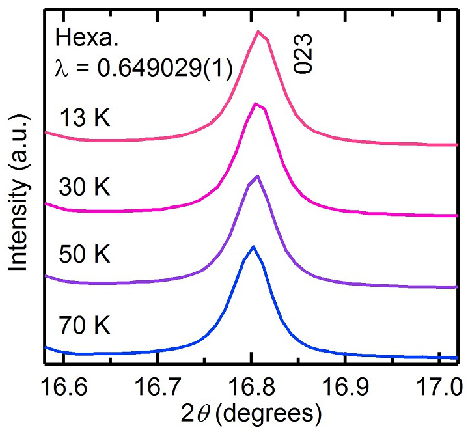}
\caption{\label{fig:FigS1}
Low temperature synchrotron x-ray powder diffraction data for hexagonal sample. No signs of broadening and splitting appear in 023 Bragg peak.}
\end{figure}

In the main text, powder synchrotron x-ray diffraction data down to 80 K are presented in Figs.2 (a) and (b). Fig.\ref{fig:FigS1} shows data in lower temperature region for hexagonal sample, collected using a cryostat in SPring-8 BL02B2. For low-temperature powder x-ray diffraction experiments, a closed-type cryostat was employed. Fig.\ref{fig:FigS1} clearly shows no signs of splitting or broadening of a Bragg peak down to 13 K, indicating the hexagonal symmetry is retained in the hexagonal sample down to the lowest temperature.

Structural parameters obtained using room temperature synchrotron single crystal x-ray diffraction data are listed in Table S1. Coordinates of Cu and Sb ions, which occupy the crystallographically equivalent sites, can be refined independently, taking advantage of a large difference in atomic form factors. Details of the experimental conditions and the obtained results are shown in Table S2. A crystal information file (CIF) of the crystal structure of Ba$_3$CuSb$_2$O$_9$ derived by the analysis can be obtained free of charge from The Cambridge Crystallographic Data Centre via www.ccdc.cam.ac.uk. CCDC 943653 and 943654 contain the supplementary crystallographic data for this paper.\\

\begin{table}[H]
\label{tab:Table1}
\begingroup
\renewcommand{\arraystretch}{1}
\caption{Crystallographic parameters of Ba$_3$CuSb$_2$O$_9$ hexagonal phases with space group $P6_3/mmc$ at 300 K for orthorhombic and hexagonal samples. The general temperature factor expression is exp(-2$\pi^2$($a^2$$U_{11}$$H^2$ + $b^2$$U_{22}$$K^2$ + $c^2$$U_{33}$$L^2$ + 2$ab$$U_{12}$$HK$ + 2$ac$$U_{13}$$HL$ + 2$bc$$U_{23}$$KL$)). The occupancy is fixed at 0.5 in Cu and Sb(2) sites.}
\scalebox{0.8}[0.8]{
\begin{tabular}{lccccccccccc}
\hline
\hline
\multicolumn{12}{c}{Ba$_{3}$CuSb$_{2}$O$_{9}$ hexagonal phase using orthorhombic sample}\\
\hline
\multicolumn{12}{l}{$a,b$ $=$ 5.81290(10) \AA, $c$ = 14.3532(6) \AA, $\alpha$ $=$ 90 $^\circ$, $\beta$ $=$ 90 $^\circ$, $\gamma$ $=$ 120 $^\circ$.} \\
\hline
site~~ & Wyck. & Occ.~~ & $x/a$ & $y/b$ & $z/c$ & $U_{11}$ & $U_{22}$ & $U_{33}$ & $U_{12}$ & $U_{13}$ & $U_{23}$\\
\hline
Ba(1)~~ & 2$b$ & 1~~   & 0 & 0 & 1/4 & 0.00987(7) & 0.00987(7) & 0.00721(7) & 0.00494(3) & 0 & 0 \\
Ba(2)~~ & 4$f$ & 1~~ & 1/3 & 2/3 & 0.09862(3) & 0.01547(7) & 0.01547(7) & 0.02416(14) & 0.00774(4) & 0 & 0 \\
O(1)~~ & 12$k$ & 1~~ & -0.1619(3) & -0.3239(6) & 0.07990(14) &0.0201(6) & 0.0196(9) & 0.0151(6) & 0.0098(4) & 0.0089(7) & 0.0045(3) \\
O(2)~~ & 6$h$ & 1~~ & 0.5158(4) & 1.0317(8) & 1/4 & 0.0169(10) & 0.0163(13) & 0.0350(18) & 0.0082(7) & 0 & 0 \\
Sb(1)~~ & 2$a$ & 1~~ & 0 & 0 & 0 & 0.00742(7) & 0.00742(7) & 0.00589(7) & 0.00371(3) & 0 & 0 \\
Sb(2)~~& 4$f$  & 0.5~~ & 1/3 & 2/3 & -0.1570(2) & 0.0073(4) & 0.0073(4) & 0.0036(2) & 0.00365(19) & 0 & 0 \\
Cu~~ & 4$f$ & 0.5~~ & 1/3 & 2/3 & -0.1541(5) & 0.0078(7) & 0.0078(7) & 0.0095(13) & 0.0039(4) & 0 & 0 \\
\hline
\hline
\\
\hline
\hline
\multicolumn{12}{c}{Ba$_{3}$CuSb$_{2}$O$_{9}$ hexagonal phase using hexagonal sample}\\
\hline
\multicolumn{12}{l}{$a,b$ $=$ 5.81260(10)  \AA, $c$ = 14.35040(10) \AA, $\alpha$ $=$ 90 $^\circ$, $\beta$ $=$ 90 $^\circ$, $\gamma$ $=$ 120 $^\circ$.} \\
\hline
site~~ & Wyck. & Occ.~~ & $x/a$ & $y/b$ & $z/c$ & $U_{11}$ & $U_{22}$ & $U_{33}$ & $U_{12}$ & $U_{13}$ & $U_{23}$\\
\hline
Ba(1) & 2$b$& 1~~ & 0 &	0 &	1/4 &	0.00996(6)	&	0.00996(6)	&	0.00680(7)	&	0.00498(3) & 0 & 0 \\
Ba(2) & 4$f$ & 1~~ & 1/3 & 2/3	&	0.09809(3)	&	0.01498(6)	&	0.01498(6)	&	0.02387(13)	&	0.00749(3) &	0	&	0	\\
O(1) & 12$k$ & 1~~ & -0.1618(2)	&	-0.3237(5)	&	0.08003(12)	&	0.0193(5)	&	0.0193(7)	&	0.0154(6)	&	0.0096(4) &	0.0085(6)	&	0.0043(3)	\\
O(2) & 6$h$ & 1~~ & 0.5164(4)	&	1.0328(7)	&	1/4	&	0.0189(9)	&	0.0167(11)	&	0.0302(14)	&	0.0084(6) &	0	&	0	\\
Sb(1) & 2$a$ & 1~~ & 0 &	0	&	0	&	0.00668(5)	&	0.00668(5)	&	0.00516(6)	&	0.00334(2) &	0	&	0	\\
Sb(2) & 4$f$ & 0.5~~ & 1/3	&	2/3	&	-0.1587(2)	&	0.00706(11)	&	0.00706(11)	&	0.0025(4)	&	0.00353(5) &	0	&	0	\\
Cu & 4$f$ & 0.5~~ &  1/3	&	2/3	&	-0.1509(4)	&	0.0069(2)	&	0.0069(2)	&	0.0021(8)	&	0.00345(11) &	0	&	0	\\
\hline
\hline
~\\
\end{tabular}
}
\endgroup
\end{table}

\begin{table}[H]
\label{tab:Table2}
\begingroup
\renewcommand{\arraystretch}{1}
\caption{Details and results of the x-ray experiments and crystal structural analysis of Ba$_3$CuSb$_2$O$_9$ hexagonal phases at 300 K for orthorhombic and hexagonal samples.}
\scalebox{0.8}[0.8]{
\begin{tabular}{lccc}
\hline
\hline
~~ & ~~ & orthorhombic sample~ & hexagonal sample~\\
\hline
Wavelength of x-ray (\AA) & ~~ & 0.42016 & 0.35451\\
Number of unique reflections & ($F_o$ $>$ 4$\sigma$($F_o$)) & 1776 & 1509 \\
Space group & ~~ & $P6_3/mmc$ & $P6_3/mmc$\\
Lattice parameters : & $a$ (\AA) & 5.81290(10) & 5.81260(10) \\
~~ & $c$ (\AA) & 14.3532(6) & 14.35040(10) \\
Residual factors : & $R$ & 0.0551 & 0.0371 \\
~~ & $R_{w}$ & 0.0915 & 0.0843 \\
Goodness-of-fit indicator : & $S$ & 1.111 & 1.067\\
\hline
\hline
\\
\end{tabular}
}
\endgroup
\end{table}

{\bf 2. Four-circle x-ray diffraction experiments}\\

\begin{figure}
\includegraphics[width=12cm]{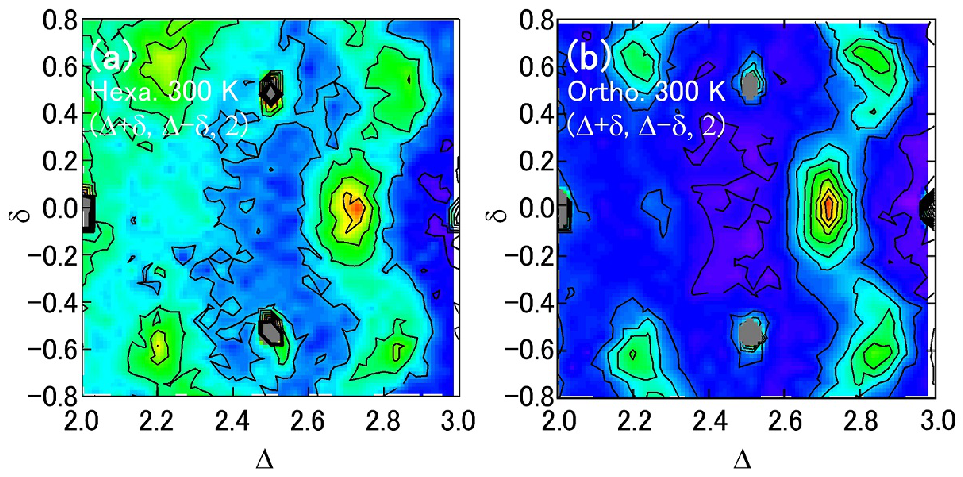}
\caption{\label{fig:FigS2}
(a)(b) Four circles x-ray diffraction experiments for hexagonal (a) and orthorhombic (b) samples at 300 K.}
\end{figure}

 In order to clarify whether there are any differences in short-range structures between the hexagonal and the orthorhombic samples, we performed diffuse x-ray scattering measurements using a point detector. Four-circle x-ray diffraction measurements were performed with a conventional Mo $K_\alpha$ x-ray generator equipped with a bent-graphite monochromator. Scattered x-rays were detected by a photon counting point detector. While the short-range structures have been already studied by observing the diffuse peaks using a photography method \cite{rf:1}, parameters such as correlation length can be estimated more accurately by the usage of a four circle diffractometer. The diffuse intensity maps around (2.5 2.5 2) are presented in Figs.\ref{fig:FigS2} (a) and (b). The overall features of the diffuse intensity distribution are very similar between two samples. Parameters of the diffuse peaks such as position, width and intensity, are comparable between them. Using the FWHM parameter obtained from the diffuse peak at around (3.33 0 2) position, the correlation length along $H$ direction is estimated to be almost comparable: 9.76 $\pm$ 0.32 \AA~ for orthorhombic sample and 10.18 $\pm$ 0.39 \AA~ for hexagonal sample, respectively. Therefore, we can conclude that the short range arrangements of the Cu and Sb in the two samples are almost the same.\\

{\bf 3. Theoretical consideration about the control parameter for tuning the ground state between the low temperature hexagonal and the orthorhombic phases}\\

The Cu-Sb coordinates for the orthorhombic and hexagonal samples apparently differ, as we discussed in the main text, whereas other structural parameters, such as lattice constants, are comparable (Table $S$I). Our four-circle x-ray diffraction measurements using single crystals show that the width and intensity of the broad superlattice peaks of the hexagonal and orthorhombic samples remain almost comparable, as shown in Fig. $S$2, which indicates that they have similar short-range structures. This situation raises the question of what is the key control parameter for tuning the ground state between the low temperature hexagonal and orthorhombic phases. In the present materials, long-range Coulomb interactions acting on the dipolar Cu-Sb dumbbells prevent the formation of both a regular triangular lattice and a long-range honeycomb lattice of the Cu-ion network. Instead, our x-ray diffraction experiments suggest that a short-range honeycomb lattice structure is established, embodied by the random replacement of Cu by Sb ions. Previous theoretical studies indicate that a uniform orbital order, which induces an orthorhombic lattice distortion, is favoured in the regular triangular lattice \cite{rf:14, rf:18}. However, the orbital liquid state, which maintains the hexagonal lattice symmetry, should be stable in the regular honeycomb lattice structure \cite{rf:15}. One possible explanation is that the present systems lie near the hexagonal-orthorhombic transition line so that a slight randomness in the Cu-Sb replacements, which is too small to be detected by our present structural measurements, would drive the system into the orthorhombic state. This would suggest that we could attain the orbital quantum phase transition by properly tuning the randomness, which would enable us to study the associated novel quantum phenomena.